\documentclass[aps,prb,twocolumn,superscriptaddress,10pt]{revtex4-2}
\usepackage[utf8]{inputenc}
\usepackage{libertine}
\usepackage[libertine]{newtxmath}
\usepackage{mathtools,graphicx,microtype,bm}
\usepackage[cal=boondoxo,frak=boondox]{mathalfa}
\usepackage[colorlinks=true,allcolors=blue]{hyperref} 

\begin{document}
\title{Quasi-bound Electron Pairs in Two-Dimensional Materials with a Mexican-Hat Dispersion}

\author{Vladimir A.\ Sablikov}
\email[E-mail:]{sablikov@gmail.com}
\author{Aleksei A.\ Sukhanov}
\email[E-mail:]{aasukhanov@yandex.ru}
 
\affiliation{Kotelnikov Institute of Radio Engineering and Electronics, Fryazino Branch, Russian Academy of Sciences, Fryazino, Moscow District, 141190, Russia}

\begin{abstract}
We study quasi-bound states of two electrons that arise in two-dimensional materials with a Mexican-hat dispersion (MHD) at an energy above its central maximum. The width of the resonance of the local density of states created by pairs is determined by the hybridization of atomic orbitals, due to which the MHD is formed. The mechanism of the quasi-bound state formation is due to the fact that effective reduced mass of electrons near the MHD top is negative. An unusual feature of quasi-bound states is that the resonance width can vanish and then they transform into bound states in continuum. We study in detail the quasi-bound states for topological insulators, when the MHD is due to the hybridization of inverted electron and hole bands. In this case, the resonance width is extremely small at weak hybridization. The highest binding energy is achieved for singlet quasi-bound pairs with zero angular number.
\end{abstract}
    
\maketitle

\section{Introduction}\label{Intro}

The band dispersion in crystals is known to play a key role in the formation of quantum states in the presence of electron-electron (e-e) interaction, in the processes of e-e scattering and, ultimately, in electron transport. In this paper we address a little-understood problem of two-electron states in two-dimensional (2D) materials with a Mexican-hat dispersion (MHD) of the band spectrum. A dispersion of this type combines two nontrivial features at once, which is obvious from the form of the MHD\@. First, these are Van Hove features on the bottom and top of the Mexican hat, widely discussed in the literature, as well as the presence of two Fermi surfaces in the energy range between the top and bottom of the hat. The second feature is that the effective mass of quasiparticles has different signs in the regions of $k$-space near the top of the Mexican hat and outside it. In addition, an inherent property of the band eigenstates is that they are really constructed from several atomic orbitals due to that the MHD is formed. Therefore the eigenstates are spinors containing not only spin components, but also pseudospin components describing atomic orbitals.  

The first feature mentioned above has been studied quite well. The van Hove singularity generates many striking effects due to e-e interaction, such as the formation of a stable ferromagnetic phase~\cite{PhysRevB.75.115425,PhysRevLett.114.236602} and a multiferroic phase appearing when lattice strains are also involved~\cite{PhysRevLett.116.206803}. Van Hove singularity significantly enhances the superconducting pairing in 2D systems~\cite{PhysRevLett.56.2732,PhysRevLett.98.167002}.

But the feature associated with the presence of negative and positive effective masses has been poorly studied. In a recent paper~\cite{SABLIKOV2023115492}, we studied the single-particle states of electrons interacting with a repulsive impurity in a MHD material and found out two main points. First, the presence of a negative effective mass leads to the appearance of quasi-bound states with the energy above the top of the MHD\@. Under certain conditions, these states can transform into bound states in a continuum (BICs). Second, the properties of quasi-bound states, such as the resonance width, the possibility of their transformation into a BIC, and the resonances of skew scattering of electrons by impurities, are correctly described only if the fact, that the wave functions are mixtures of atomic orbitals involved in the formation of MHD, is taken into account. Moreover, it is important that the weight with which each the atomic orbitals enters the total wave function, that is, the envelopes of the wave functions corresponding to these orbitals, are determined by the equations of motion. Therefore, a correct description of quasi-bound states and, in general, the electron dynamics in an external field should be carried out beyond the limits of the single-band approach and based on more realistic models. 

\begin{figure}
    \centerline{\includegraphics[width=0.9\linewidth]{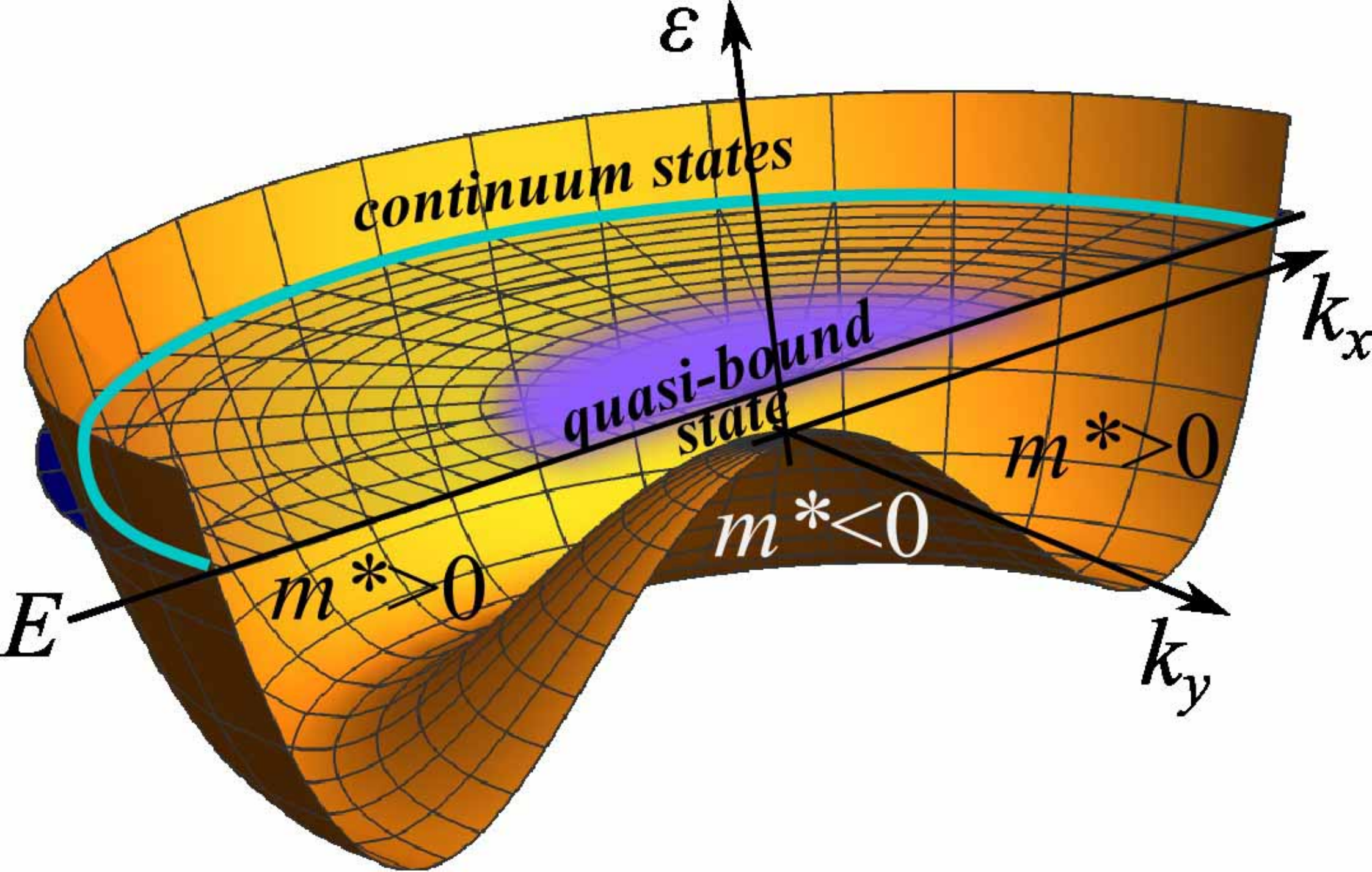}}
    \caption{Schematic view of energy diagram of the MHD and a quasi-bound state at the energy $E$ in the $k$ space. $m^*$ is the reduced effective mass formed due to mixing of different atomic orbitals. The plane section at level $E$ is a set of wave vectors, on which the quasi-bound state is formed. Purple shows the region due to which the localized component of the quasi-bound state arises.}
   \label{fig1} 
\end{figure}

The single-band approach based on a simplified Hamiltonian of the form
\begin{equation}
    H= a_1 k^4 - a_2 k^2 + E_0\,
    \label{eq.1_single-band}
\end{equation}
can be justified only for sufficiently low-energy perturbations, though it catches the van Hove singularity. The point is that a perturbing potential of an impurity or e-e interaction, significantly affects the hybridization of atomic orbitals that form the quantum states. More formally, the resulting states cannot be represented in a single-band basis. Direct calculation of single-particle quasi-bound states on a point defect, not included in Ref.~\cite{SABLIKOV2023115492}, leads to results that differ significantly from the two-band model. In particular, the single-band approach misses the possibility of the transformation of a quasi-bound state into a BIC\@.

Based on these results for single-particle states, one can expect that interacting electrons also form quasi-bound two-particle states with energies above the MHD top. The mechanism of the quasi-bound state formation can be explained as follows. The interaction potential perturbs the states of free electrons so that resulting wave function is expanded over a set of single-particle states in the entire $k$-space as illustrated in Fig.~\ref{fig1}. The components from the region of small $k$ near the top of the hat are characterized by a negative effective mass. The reduced effective mass of two electrons in such states is negative and therefore they are attracted to each other. The components from the region of large $k$ with positive mass are repelled. Thus, the complete two-particle wave function is partially localized near the center of mass of the pair, and partially propagates to infinity, merging with the continuum of band states. It is important that the localized component and the continuum states are created mainly by different atomic orbitals, and therefore overlap weakly.

Of fundamental importance is the question of the reduced mass. The concept often used in the one-band approximation that it is determined by a dispersion equation of type Eq.~(\ref{eq.1_single-band}) can be justified only when the energy deviates little from the value determined by the band dispersion. This is obviously not the case when the profile of the MHD is quite well pronounced and the energy difference between the top and the bottom of the MHD is large enough. This question was studied in more detail in Ref.~\cite{PhysRevB.95.085417} for bound electron pairs with energy in the gap. It was shown that the effective reduced mass in the two-band model is very different from that obtained from the single-band dispersion. 

A correct study of two-electron quasi-bound states should be based on a specific Hamiltonian of the MHD material. The MHD appears in a variety of materials described by Hamiltonian of very different structure, such as topological insulators in which the MHD naturally appears as a consequence of the band inversion~\cite{PhysRevB.85.161101}, bilayer graphene~\cite{mccann2007low,doi:10.1126/sciadv.aau0059}, monolayers of group III-IV chalcogenides~\cite{PhysRevB.95.115409,doi:10.1063/1.4928559,C5NR08982D}. But the most general and universal model of the MHD formation is a hybridization of inverted electron-like and hole-like subbands. In this case, of particular interest is the situation when the hybridization of these subbands depends on the spin and momentum of the electron, since this hybridization gives rise to a spin-orbit coupling and it can be expected that the resulting states will have nontrivial properties such as helicity. This situation is described by the Bernevig-Hughes-Zhang (BHZ) model~\cite{BHZ}. It is realized in many 2D topological insulators such as HgTe/CdHgTe~\cite{doi:10.1126/science.1148047} and InAs/GaSb~\cite{knez2012quantum}. Note that quasi-bound states for graphene were considered in Ref.~\cite{kochelap2022rotating} on the basis of the single-band approximation, but the validity of its applicability remains questionable. 

In this paper we present a consistent theory of quasi-bound states of two electrons with repulsive interaction for 2D materials described by the BHZ model, which allows one to study properties of the quasi-bond states in realistic systems. These properties, as it turned out, are largely determined by the hybridization parameter of the electron and hole bands, which also determines the conditions under which the MHDs appear. We find the energy and width of the resonance of local density of states arising due to the quasi-bound state and study their dependence on the hybridization parameter $a$ and the angular quantum number. We explore how the resonances depend on the amplitude and radius of the e-e interaction potential in a simplified model. Most intriguing finding is that the resonance width is unexpectedly small, since it is proportional to $a^4$. The angular motion of the electrons is not crucially important for the appearance of the quasi-bound state and their spin, though affects the resonances. In addition, we show that under certain condition, the quasi-bound state can transform into a BIC and trace the evolution of this transformation. 

\section{Model and basic equations}\label{S:model}
Within the frame of the BHZ model~\cite{BHZ}, the MHD arises in the topological phase due to the hybridization of the inverted electron and hole bands. Single-particle states are built on a four-band basis of electron and hole Bloch orbitals: $\Phi=(|e\uparrow\rangle,|h\uparrow\rangle,|e\downarrow\rangle,h\downarrow\rangle)^T$. The hybridization of electron and hole bands is described by the parameter $A$. For simplicity, we will consider the situation when the electron and hole bands are symmetric, and the $S_z$ symmetry is not violated. In this case, the Hamiltonian reads~\cite{BHZ}
\begin{align}
  \hat{H}_0(\hat{\mathbf{k}})&=
  \begin{pmatrix}
   \hat{h}_{\mathbf{k}} &0 \\
   0 & \hat{h}^*_{-\mathbf{k}}\\
  \end{pmatrix},\nonumber \\
 \hat{h}_{\mathbf{k}}=&\,\mathbf{d}_{\mathbf{k}}\!\cdot\!\hat{\bm{\tau}},\quad \mathbf{d}_{\mathbf{k}}=(ak_{x},-ak_{y},-1+k^2)\,,
\end{align}
where $\hat{\bm{\tau}}$ is the vector of Pauli matrices. The band dispersion has the following form~\cite{franz2013topological}
\begin{equation}\label{eq:band_dispersion}
 \varepsilon(k)=\pm\sqrt{1+(a^2-2)k^2+k^4}\,.   
\end{equation}
Here the energy $\varepsilon$ is normalized to the mass term $|M|$, the wave vector $\mathbf{k}$ is normalized to $\sqrt{|M/B|}$, with $B$ being the parameter of electron and hole band dispersion, and $a$ is the dimensionless hybridization parameter $a=A/\sqrt{|MB|}$, which largely determines the properties on the quasi-bound states.

The MHD is seen to arise at sufficiently weak hybridization, when $a<\sqrt{2}$. The MHD top lies at $\varepsilon=1$, while the bottom is at $\Delta=|a|\sqrt{1-a^2/4}$.  For definiteness, in what follows we consider the states in the conduction band. The single-particle states with the wave vector in the region near the MHD top consist mainly of the hole-like band states and have a negative effective mass. The states with $k$ outside this region are composed mainly by the electron-like states with positive effective mass. The relative motion of two electrons and the possibility that they form a quasi-bound state are determined by the contributions of all combinations of all admissible single-particle states of both electrons. Therefore, the complex problem of the spectrum and structure of the quasi-bound states of a pair must be solved on the basis of two-particle equations.

 The two-particle wave functions are built on the two-particle basis $\Phi(1)\otimes \Phi(2)$. We study two-particle states based on the approach previously developed for the BHZ model in Ref.~\cite{PhysRevB.95.085417}. The two-electron Hamiltonian reads
\begin{equation}\label{H1}
    \hat{H}(1,2)=\hat{H}_0(\mathbf{\hat{k}}_1)\oplus\hat{H}_0(\mathbf{\hat{k}}_2)
    +v(|\mathbf{r}_1-\mathbf{r}_2|)\cdot\hat{\mathbf{I}}_{16\times 16}\,,
\end{equation} 
where $\hat{H}_0(\mathbf{\hat{k}}_{1,2})$ is the single-particle Hamiltonian of the BHZ model, $v(|\mathbf{r}_1-\mathbf{r}_2|)$ is the interaction potential. 

Because of the $S_z$-symmetry of the system, there are three types of two-particle states with well-defined total spin $S_z=0$ and $S_z=\pm 1$. One state is a singlet with opposite electron spins, and the other two are triplets with parallel spins. The singlet state is described by four independent envelope functions and each of two triplet states is described by three envelope functions. 

The envelope functions are determined by the corresponding Schr\"odinger equations, which, however, are rather complicated in general form due to the fact that the relative motion of the particles is not separated from the motion of the center of mass. We consider here the most interesting situation when the total momentum of the electron pair is zero. Then, the equations for the envelope functions are simplified and can be solved analytically for specific form of the e-e interaction potential. The motion of a pair as a whole can be studied perturbatively~\cite{SABLIKOV2022127956}. 

In the case of the singlet states, the envelope functions $\psi_3(\mathbf{r})$, $\psi_4(\mathbf{r})$, $\psi_7(\mathbf{r})$, $\psi_8(\mathbf{r})$, with $\mathbf{r}$ being vector of relative position of electrons, are defined by the following equation system~\cite{PhysRevB.95.085417} 
\begin{equation}\label{eq:singlet_eqs}
\left\{\!\!
 \begin{array}{rl}
  \left[v(r)-\varepsilon-1+\hat{k}^2\right]\psi_3(\mathbf{r})+a\hat{k}_-\psi_4(\mathbf{r})+a\hat{k}_+\psi_7(\mathbf{r}) &=0 \\
  a\hat{k}_+\psi_3(\mathbf{r})+\left[v(r)-\varepsilon\right]\psi_4(\mathbf{r})+a\hat{k}_+ \psi_8(\mathbf{r}) &=0\\
  a\hat{k}_-\psi_3(\mathbf{r})+\left[v(r)-\varepsilon\right]\psi_7(\mathbf{r})+a\hat{k}_- \psi_8(\mathbf{r}) &=0\\
  a\hat{k}_-\psi_4(\mathbf{r})+a\hat{k}_+\psi_7(\mathbf{r})+\left[v(r)-\varepsilon+ 1-\hat{k}^2\right]\psi_8(\mathbf{r}) &=0\,,
 \end{array}\right.
\end{equation}
where $\varepsilon$ is the two-electron energy per one electron, i.e.\ the energy of two electrons divided by 2.  

Since the system has the circular symmetry, the wave functions are expanded over angular harmonics
\begin{equation}
  \psi_i(\mathbf{r})=\sum\limits_m \psi_{i,m}(r)e^{i m \phi}\,,
\end{equation}
with angular quantum number $m=0, \pm 1, \pm 2,\dots$
 
The wave function of singlet states, which satisfies the requirement of antisymmetry with respect to the permutation of particles, is expressed in terms of the functions $\psi_i$ as follows
\begin{equation}\label{eq:singlet_WF}
  \begin{split}
   \Psi_m^{(S)}(1,2)&= C\left[\psi_{3m}(r)\left(|e\uparrow e\downarrow\rangle-|e\downarrow e\uparrow\rangle\right)\right.\\
   &\left.+ \psi_{4m}(r)e^{i\varphi}\left(|e\uparrow h\downarrow\rangle+|h\downarrow e\uparrow\rangle\right)\right.\\
   &\left.+ \psi_{7m}(r)e^{-i\varphi}\left(|h\uparrow e\downarrow\rangle+|e\downarrow h\uparrow\rangle\right)\right.\\
   &\left.+ \psi_{8m}(r)\left(|h\uparrow h\downarrow\rangle-|h\downarrow h\uparrow\rangle\right)\right].
  \end{split}
\end{equation}
It is seen, this wave function can not be factorized into orbital and spin functions as in usual case of spin-independent Hamiltonian. Important conclusion from Eq.~(\ref{eq:singlet_WF}) is that the angular quantum number can be only even, $m=0, \pm 2,\dots$. Otherwise, the $\Psi_m^{(S)}(1,2)$ is not antisymmetric with respect to the electron permutation. 

The triplet states are studied in a similar way. The essential differences are as follows. There are two types of triplet states, in which the spins of both electrons are directed up or down. In both cases, the states are described by three independent envelope functions, since two of four envelope functions are related by the simple relation $\psi_5(\mathbf{r})=-\psi_2(\mathbf{r})$.  So, the spin-up triplet is described by $\psi_1(\mathbf{r})$, $\psi_2(\mathbf{r})$, and $\psi_6(\mathbf{r})$.  They are defined by equations similar to Eq.~(\ref{eq:singlet_eqs}). The wave function antisymmetric with respect to particle permutation has the form:
\begin{equation}\label{eq:triplet_WF}
  \begin{split}
   \Psi_m^{(T\uparrow)}(1,2)= C&\left[\psi_{1m}(r)e^{i\varphi}|e\uparrow e\uparrow\rangle\right.\\
   &\left.+ \psi_{2m}(r)\left(|e\uparrow h\uparrow\rangle-|h\uparrow e\uparrow\rangle\right)\right.\\
   &\left.+ \psi_{6m}(r)e^{-i\varphi}|h\uparrow h\uparrow\rangle\right].
  \end{split}
\end{equation}
In this case, again, the angular quantum number $m$ can be only even. 

\section{Quasi-bound states in weakly hybridized electron and hole bands}\label{S:weak_hybrid}%
The effects of the nonmonotonic shape of the MHD are expected to be most significant when the top and bottom of the MHD are strongly separated in energy. In the BHZ model, this occurs when the electron and hole bands are weakly coupled and the hybridization parameter is small, $a\ll 1$. This situation is most interesting because it allows one to reveal the main features of quasi-bound states within a relatively simple approach.  

First of all, we note that in the case of $a=0$, the subsystems of electron- and hole-like states are decoupled. So, in Eq.~(\ref{eq:singlet_eqs}) for singlet states, the functions $\psi_4$ and $\psi_7$ turn to zero and the equation system splits into two independent equations for the functions $\psi_3$ and $\psi_8$. The first equation for $\psi_3$ describes a particle with a positive mass in the repulsive potential. Thus, the function $\psi_3$ describes propagating states with a continuous spectrum. The last equation, on the contrary, describes a particle with a negative mass. Therefore, its spectrum is discrete in the energy range $1<\varepsilon <1+\max(v(r))$, and the function $\psi_8$ is localized. When $a\neq 0$, these states hybridize, resulting in resonant states.
 
A constructive approach to solving the system of Eq.~(\ref{eq:singlet_eqs}) is to consider it as two weakly coupled subsystems of electron-like and hole-like states, and treat the coupling between them perturbatively, as in the Fano-Anderson model~\cite{PhysRev.124.1866,mahan2013many}.

To do this, we eliminate the functions $\psi_{4}$ and $\psi_7$ from Eq.~(\ref{eq:singlet_eqs}) and, assuming that $v(r) \ll 1$, pass to a system of two equations for $\psi_{3}$ and $\psi_{8}$. In addition, we take into account that the energy interval of interest to us (where the resonance can occur) is narrow $1<\varepsilon<1+\max{[v(r)]}$. In this way, Eq.~(\ref{eq:singlet_eqs}) takes the form 
\begin{equation}\label{Shred_eff}
\hat{\mathcal{H}} \Psi= \delta\varepsilon\cdot \Psi \,,
\end{equation} 
where $\Psi =(\psi_3, \psi_8)^T$, $\delta\varepsilon=\varepsilon-1$. The Hamiltonian $\hat{\mathcal{H}}$ is
\begin{equation}
  \hat{\mathcal{H}}=\hat{\mathcal{H}}_0+\hat{\mathcal{H}}_a\,,  
\label{F-A_hamiltonian}
\end{equation}
with
\begin{equation}
 \hat{\mathcal{H}}_0=
 \begin{pmatrix}
 \mathbf{\hat{k}}^2\!-2\! +\! v(r) \! &\! 0\\
  \! 0 & \! -\mathbf{\hat{k}}^2 +\! v(r)\\
\end{pmatrix},\;
\end{equation}
\begin{equation}
\hat{\mathcal{H}}_a = a^2 \hat{\beta}
\begin{pmatrix}
   1 & 1\\
   1 & 1
  \end{pmatrix}.
\end{equation}
where  
$ \hat{\beta}\!=2\left[\hat{k}^2\!+\frac{dv(r)}{dr}\frac{d}{dr}\right]. $

When $a=0$, the Schr\"odinger equation~(\ref{Shred_eff}) reduces to two simple independent equations 
\begin{equation}\label{psi30}
 \left[v(r)-\delta\varepsilon-2+\hat{k}^2\right]\psi_{3}^{(0)}(r)=0 ,
\end{equation}
\begin{equation}\label{psi80}
 \left[v(r)-\delta\varepsilon-\hat{k}^2\right]\psi_{8}^{(0)}(r)=0 .
\end{equation}
Equation~(\ref{psi30}) describes the continuous spectrum $\delta\varepsilon_k=-2+k^2$ of two-particle states with the wave functions of the form $\psi_3^{(0)}=\psi_{k,m}(r)e^{im\varphi}$, with $k$ being momentum and $m$ angular quantum number. Equation~(\ref{psi80}) has two types of eigenfunctions. One type states has a discrete spectrum $\delta\varepsilon_{n,m}$ at positive energy. They are described by wave function $\psi_8^{(0)}=\phi_{n,m}(r)e^{im\varphi}$ localized in the region of the e-e interaction. Other states have a continuous spectrum $\delta\varepsilon_k=-k^2$ and are described by propagating wave functions $\psi_8^{(0)}=\phi_{k,m}(r)e^{im\varphi}$.

When $a\neq0$, the hybridization of discrete  localized states  with states of the continuous spectrum leads to a shift and, what is especially important, to a broadening of the discrete levels, i.e.\ to the transformation of the bound state into a resonant one.
In the case of $a\ll 1$, the wave function of such states can be constructed from the wave functions of the propagating continuum states $\psi_{k,m}e^{im\varphi}$ and localized states $\phi_{n,m}e^{im\varphi}$ similarly to the theory of quasi-bound states in the repulsive impurity potential~\cite{SABLIKOV2023115492}. 

For simplicity, we confine ourselves to considering only one localized state with energy $\delta\varepsilon^{(0)}_0$ and a wave function $\phi$ with zero angular quantum number. In addition, we neglect the hybridization of the continuum of the $\psi_{k,m}e^{im\varphi}$ states with the continuum of $\phi_{k,m}(r)e^{im\varphi}$ states at $\delta\varepsilon < 0$. The wave function satisfying the Eq.~(\ref{Shred_eff}) is represented as
\begin{equation}\label{Psi2_F}
 \Psi=\left(\sum\limits_{k,m}A_{k,m} \psi_{k, m}e^{im\varphi}, B\phi\right)^T.
\end{equation} 
Though the momentum is not conserved in this state, it is convenient to characterize the wave function by the wave vector $\mathcal{K}=\sqrt{2+\delta\varepsilon}$ corresponding to the wave vector that would be without the e-e interaction.

Following the standard procedure of the Fano-Anderson model~\cite{PhysRev.124.1866,mahan2013many}, we arrive at the following eigenfunctions:
\begin{equation}\label{Psi2_K}
 \Psi_\mathcal{K}
 =B_\mathcal{K}
 [\phi+a^2 Z_\mathcal{K}\langle \mathcal{K},0| \hat{\beta}|\phi\rangle \psi_{\mathcal{K},0}^{(0)}+a^2\sum_{k}\mathcal{P}\frac{\langle k,0|\hat{\beta}|\phi\rangle}{\mathcal{K}^2-k^2}\psi_{\mathcal{K},0}^{(0)}]\,,
\end{equation} 
where $\mathcal{P}$ denotes the principal value, all angular components except $m=0$ vanish because we consider the localized state with $m=0$. The value $Z_{\mathcal{K}}$ is defined as
\begin{equation}\label{Z}
 Z_\mathcal{K}=\frac{\mathcal{K}^2-2-\delta\varepsilon^{(0)}_0+a^2\langle \phi|\hat{\beta}|\phi\rangle-\Sigma_\mathcal{K}}{a^4|\langle \mathcal{K},0|\hat{\beta}|\phi\rangle |^2 }.
\end{equation} 
The wave function amplitude $\mathit{B}_\mathcal{K}$ is
\begin{equation}\label{BK}
\mathit{B}_\mathcal{K}=\!\frac{a^2|\langle \mathcal{K},0|\hat{\beta}|\phi\rangle |}{\sqrt{\left(\mathcal{K}^2-2-\delta\varepsilon^{(0)}_0 - a^2\langle \phi|\hat{\beta}|\phi\rangle-\Sigma_\mathcal{K}\right)^2+\gamma_\mathcal{K}^2}} ,
\end{equation}
It is seen that $\mathit{B}_\mathcal{K}$ describes a resonance at the energy shifted relative the localized state energy $\delta\varepsilon^{(0)}_0$
\begin{equation}\label{Emax}
  \delta\varepsilon_0=\delta\varepsilon^{(0)}_0+a^2\langle \phi|\hat{\beta}|\phi\rangle+\Sigma_\mathcal{K}\,, 
\end{equation}
with $\Sigma_\mathcal{K}$ being the self-energy function
\begin{equation}\label{Sigm}
  \Sigma_\mathcal{K}=a^4\sum\limits_{k}\mathcal{P}\frac{1}{\mathcal{K}^2-k^2} |\langle k,0|\hat{\beta}|\phi\rangle|^2. 
\end{equation}
The resonance width is  
\begin{equation}\label{G}
  \gamma_\mathcal{K}=a^4\frac{L}{2v_K}|\langle \mathcal{K},0|\hat{\beta}|\phi\rangle|^2\,,
\end{equation}
where $L$ is the normalization length and $v_{\mathcal{K}}$ is the group velocity.

Thus, as can be seen from Eqs.~(\ref{Emax}) and~(\ref{G}) the resonance energy increases quadratically with $a$, and the resonance width is proportional to $\sim a^4$. Thus, it can be expected that the quality factor of the resonance will be extremely high at $a\ll 1$ when the MHD profile is strongly pronounced. It is also worth noting that in the limit $a\to 0$ the quasi-bound states turn into the bound states in the continuum (BICs)\@. 

Another interesting conclusion from the equation~(\ref{G}) is that, at a finite $a$, a quasi-bound state can transform into a BIC under a certain requirement on the interaction potential. This occurs when the matrix element $\langle \mathcal{K},0|\hat{\beta}|\phi\rangle$ equals zero, which is possible for some amplitude and the shape of the e-e interaction potential.

In the case considered above, the quasi-bound state appears with the zero angular quantum number. However, it can be seen from the derivation of Eqs.~(\ref{Psi2_F})~--~(\ref{G}) that quasi-bound states with $m\neq 0$ can also form. To see this, it suffices to include localized states with $m\neq 0$ in the expression for the localized state wave function $\phi_{n,m}\exp(im\varphi)$. 

Qualitatively similar results are obtained for triplet two-electron states.

\section{Quasi-bound states at a step-like interaction potential}\label{S:step}

When the e-e interaction potential is not small, the quasi-bound states can be studied for a specific form of the e-e interaction potential. In this section, we carry out such calculations for a step-like potential, which is a widely used model of a short-range interaction. As a result we find the spectrum and wave functions, as well as study their dependence on the magnitude of the potential and the radius of interaction.

The stepwise form of the e-e interaction potential $v(r)=v_0\Theta(r_0-r)$, where $v_0$ and $r_0$ are the amplitude and the radius of the interaction, allows one to solve exactly the equations for the spinor components of the wave function in each region of space where the potential does not depend on the coordinate $r$. Then, matching these solutions at the boundary $r=r_0$ we find the wave function in all space. A detailed description of this approach is given in Ref.~\cite{PhysRevB.95.085417}. Here, we present only the main results.

First, it is convenient to pass in the system of Eqs.~(\ref{eq:singlet_eqs}) to polar coordinates and expand the wave function over angular harmonics. In the case of the singlet two-particle states the spinor components of the wave function are:
\begin{equation}\label{psi-m}
\begin{array}{ll}
\psi_{3}(\mathbf{r})=\sum\limits_m \psi_{3m}(r) e^{im\varphi},\\
\psi_{4}(\mathbf{r})=\sum\limits_m \psi_{4m}(r) e^{i(m+1)\varphi}, \\
\psi_{7}(\mathbf{r})=\sum\limits_m \psi_{7m}(r) e^{i(m-1)\varphi}, \\ 
\psi_{8}(\mathbf{r})=\sum\limits_m \psi_{8m}(r) e^{im\varphi}\,.
\end{array}
\end{equation} 
 
The functions $\psi_{3m}(r)$, $\psi_{4m}(r)$, $\psi_{7m}(r)$ and $\psi_{8m}(r)$ are determined from Eqs.~(\ref{eq:singlet_eqs}) after substituting Eqs.~(\ref{psi-m}) into them. Resulting equations are solved in terms of the Bessel functions in the regions $r<r_0$ and $r>r_0$. For the energy range $\varepsilon>1$ of interest to us the following equations are obtained.

In the interaction region, $r<r_0$,
\begin{equation}\label{wave_func-}
 \begin{array}{ll}
 \psi_{3m}=& A_{1} J_m(k_+r)+A_{2} J_m(k_-r),\\
 \psi_{4m}=& A_{1}\mathcal{B}_+J_{m+1}(k_+r)+A_{2}\mathcal{B}_-J_{m+1}(k_-r),\\
 \psi_{7m}=& A_{1}\mathcal{C}_+J_{m-1}(k_+r)+A_{2}\mathcal{C}_-J_{m-1}(k_-r),\\
 \psi_{8m}=& A_{1}\mathcal{D}_+J_m(k_+r)+A_{2}\mathcal{D}_-J_m(k_-r),
 \end{array}
\end{equation} 
where
\begin{equation}\label{k}
 k_{\pm}=\sqrt{1-\frac{a^2}{2}\pm\sqrt{a^2\left(\frac{a^2}{4}-1\right)+(\varepsilon-v_0)^2}}.
\end{equation} 

In the outer region, $r>r_0$,
\begin{equation}\label{wave_func+}
 \begin{array}{ll}
\psi_{3m}=A_{3} H^{(1)}_m( k_p r)\!+\!A_{4} H^{(2)}_m(k_pr)\!+\!A_{5} K_m(\varkappa_n r)\\
\psi_{4m}=A_{3}\mathcal{K}_p H^{(1)}_{m+1}(k_pr)\!+\!A_{4}\mathcal{K}_p H^{(2)}_{m+1}(k_p r)\!+\!A_5\mathcal{K}_n K_{m+1}(\varkappa_n r)\\
\psi_{7m}=A_{3}\mathcal{L}_p H^{(1)}_{m-1}(k_p r)\!+\!A_{4}\mathcal{L}_p H^{(2)}_{m-1}(k_p r)\!+\!A_5\mathcal{L}_n K_{m-1}(\varkappa_nr)\\
\psi_{8m}=A_{3}\mathcal{M}_p H^{(1)}_{m}(k_p r)\!+\!A_{4}\mathcal{M}_p H^{(2)}_{m-1}(k_p r)\!+\!A_5\mathcal{M}_n K_{m}(\varkappa_n r)\,, 
 \end{array}
\end{equation} 
where $H^{(1, 2)}_m$ and $K_m$ are the Hankel and MacDonald functions, the wave numbers $k_p$ and $\varkappa_n$ are defined respectively as   $k_p=k_+$ at $v_0=0$, and $\varkappa_n=-ik_-$ at $v_0=0$. The coefficients $\mathcal{B}_{\pm}$, $\mathcal{C}_{\pm}$, $\mathcal{D}_{\pm}$, $\mathcal{K}_{p,n}$, $\mathcal{L}_{p,n}$, $\mathcal{M}_{p,n}$ are found by substituting Eqs.~(\ref{wave_func-}) and~(\ref{wave_func+}) into Eqs.~(\ref{psi-m}) and~(\ref{eq:singlet_eqs}). Explicit expressions are rather cumbersome, so we omit them.

The basic role is played by the coefficients $A_1, A_2, A_3, A_4$ and $A_5$, that determine the amplitudes of spatial harmonics. The coefficients $A_3, A_4$ are amplitudes of propagating waves falling on the center and outgoing, the coefficient $A_5$ is the amplitude of an evanescent wave around the interaction region. The coefficients $A_1, A_2$ are amplitudes of two undamped modes in the interaction region that interfere with each other. All five coefficients are defined from the matching of the wave functions at the boundary of the outer region and interaction region, $r=r_0$, and the normalization equation for the total function.

The matching conditions are obtained by integrating Eqs.~(\ref{eq:singlet_eqs}) over the transition region, $|r-r_0|<\delta$, between the outer and interaction regions and taking a limit $\delta \to 0$,
\begin{eqnarray}
  \psi_{3m}\bigm|_-^+ & =0\\
  \psi_{8m}\bigm|_-^+ & =0\\
  2 \frac{d\psi_{3m}}{dr}+ia(\psi_{4m}+\psi_{7m})\biggm|_-^+ & =0\\
  \frac{d\psi_{3m}}{dr}+\frac{d\psi_{8m}}{dr}\biggm|_-^+ & =0\,.
\end{eqnarray}

To find out the quasi-bound states, we consider the probability of finding the electrons in the interaction region. This probability can be estimated from the electron density integrated over the interaction region, $\langle N\rangle$. We have studied the dependence of this quantity on energy. The result presented in Fig.~\ref{fig_n(E,a)} clearly shows that $\langle N\rangle$ has a sharp maximum at a certain energy $\varepsilon_0$, which increases with increasing the hybridization parameter $a$. More detailed calculations show that for $a\ll 1$ the shift of the resonance energy is proportional to $a^2$. 

\begin{figure}
\centerline{\includegraphics[width=0.9\linewidth]{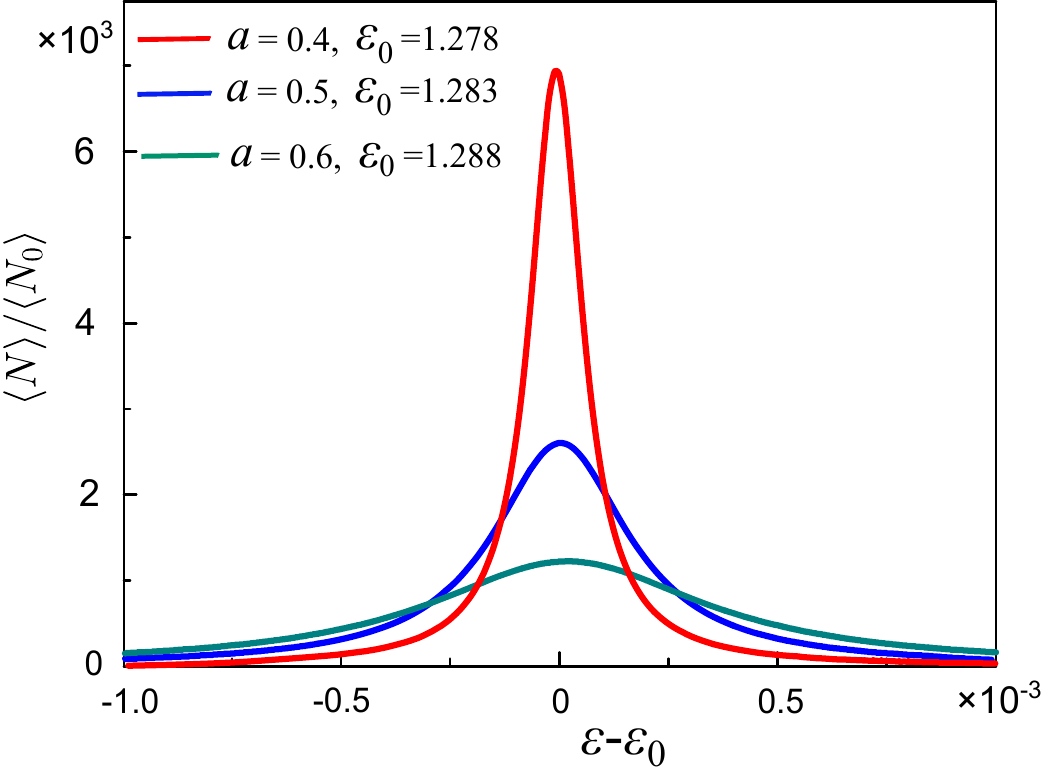}}
\caption{Resonance of the electron density $\langle N \rangle$ integrated over the interaction region for a variety of the  hybridization parameters $a=0.4,  0.5,  0.6$. $\langle N \rangle$ is normalized to its value at $v_0\to 0$, which is denoted as $\langle N_0 \rangle$. Other parameters are: $v_0=0.4$, $r_0=5$, $m=0$.}
\label{fig_n(E,a)}
\end{figure}

But the main effect of changing the hybridization parameter is a strong increase in the width of the resonance with the parameter $a$. The resonance width, $\gamma$, changes as $a^4$, which agrees with the analytical estimates of Sec.~\ref{S:weak_hybrid}, Eq.~(\ref{G}). Numerically, for $a=0.4$, the value of $\gamma$ normalized to $|M|$  is estimated as $2\cdot 10^{-4}$, and as $a$ decreases to $0.1$, it drops to $\gamma = 10^{-6}$.

The resonance energy $\varepsilon_0$ and width of the resonance $\gamma$ depend on the amplitude and radius of the e-e interaction potential, $v_0$ and $r_0$. The energy $\varepsilon_0$, as a function of $v_0$ and $r_0$, has no striking features and qualitatively corresponds to the behavior of the levels in the quantum well created by the interaction potential for hole-like states.

\begin{figure}
  \centerline{\includegraphics[width=0.95\linewidth]{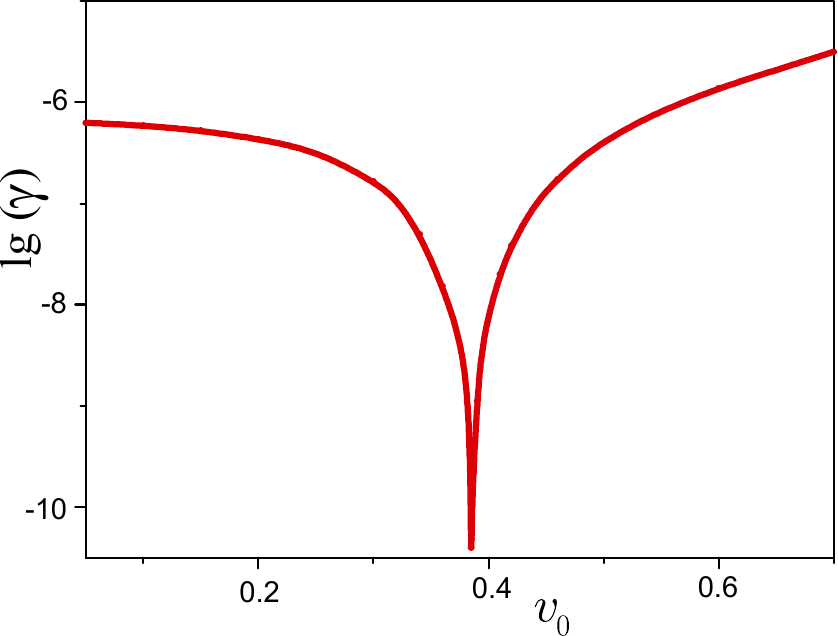}}
    \caption{Dependence of the resonance width on the amplitude of the e-e interaction potential at a given interaction radius $r_0=3.6$. The parameters used in calculation are $a=0.2$ and $m=0$.} 
   \label{fig:3}
\end{figure}

The width of the resonance is of the greatest interest, since it largely determines the lifetime of a quasi-bound pair relative to the decay into separately moving particles. Therefore, we studied in more detail how it changes depending on the parameters of the model interaction potential, which can vary over a wide range, due to the inevitable screening present in real systems. The dependence of the resonance width on $v_0$  at a fixed $r_0$ is shown in Fig.~{\ref{fig:3}}. It is clear that in a wide range of $v_0$ the resonance width is very small, and near a certain value of $v_0$,  the width $\gamma$ drops catastrophically. Accordingly, the amplitude of the resonance increases approximately the same. In the specific example in Fig.~{\ref{fig:3}}, $\gamma$ drops by a factor of $10^{-4}$. The minimum value of $\gamma$ strongly depends on $r_0$. Dependences of a similar kind also occur when $r_0$ changes for a given $v_0$.

\begin{figure}[h]
  \centerline{\includegraphics[width=0.95\linewidth]{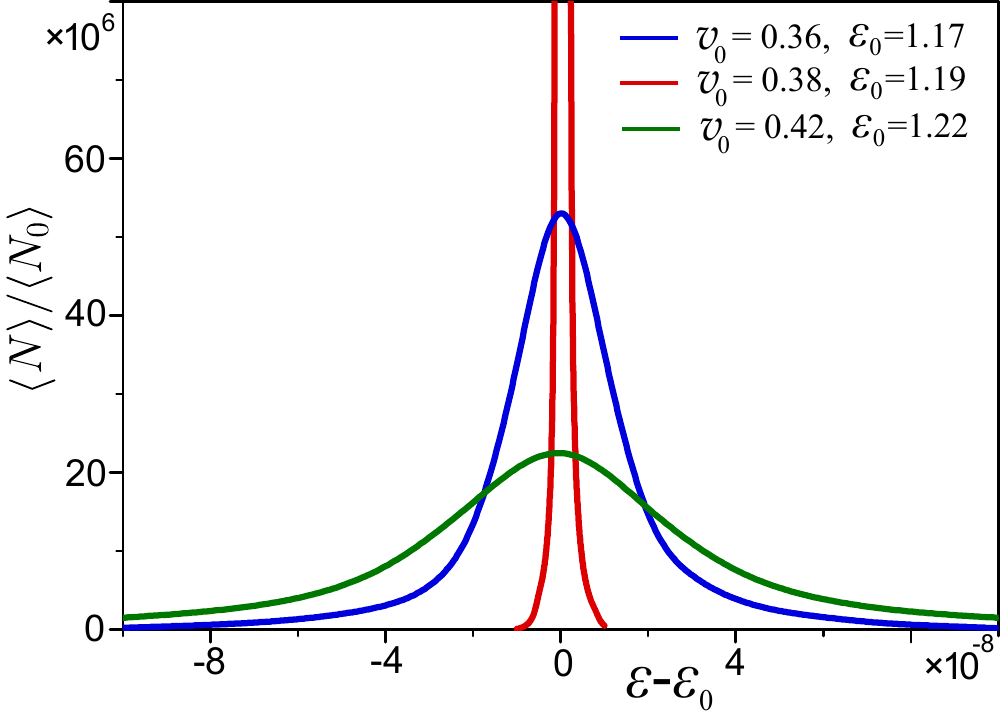}}
   \caption{Evolution of the quasi-bound state resonance with varying the amplitude of the e-e interaction potential near the singular point at a given interaction radius $r_0=3.6$ close to $r_c$. Calculations are carried out for $a=0.2$ and $m=0$.} 
   \label{fig:4}
\end{figure}

An interesting feature of this dependence is that $\gamma$ vanishes at certain values of the pair of parameters $v_0=v_c$ and $r_0=r_c$. These are singular points on the parameter plane $(v_0, r_0)$, where the BIC is formed. Near these points, $\gamma$ as a function of $v_0$ and $r_0$ has a sharp minimum, which is a precursor of the BIC. The evolution of the resonance in the vicinity of the singular point is shown in Fig.~{\ref{fig:4}} as $v_0$ changes near $v_c\approx 0.391$ and $r_0$ is fixed near $r_c \approx 3,607$. 

These results show that the width of the quasi-bound state resonance is very small and is determined by the hybridization parameter, $\gamma\sim a^4$. This means that the quasi-bound pairs are long-lived with respect to decay into individual particles. The fact that $\gamma$ is small in a wide range of the interaction potential parameters shows that this conclusion is stable with respect to the choice of parameters that determine the shape of the short-range interaction potential. Another conclusion is that there are singular points on the parameter plane $(v_0, r_0)$ at which BICs are formed. Near these points, the resonance width drops sharply. This conclusion is much more sensitive to the parameters of the interaction potential.

The step-like potential model make it possible to better understand the mechanism of the BIC formation and the conditions for the occurrence of such a state. Equations~(\ref{wave_func+}) clearly show that the two-particle state is localized in space and splits off from the continuum when the coefficients $A_3$ and $A_4$, which describe the propagating components of the wave function, vanish. Formally, $A_3$ and $A_4$ are functions of the energy $\varepsilon$ and parameters of the system, primarily $v_0$ and $r_0$. Thus, the condition for the formation of a BIC is determined by the system of equations $A_3(\varepsilon, v_0, r_0)=0$ and $A_4(\varepsilon, v_0, r_0)=0$. It is obvious that these equations are compatible only for certain values of $v_0$ and $r_0$. On the other hand, it follows from Eqs.~(\ref{wave_func-}) that a situation of a standing wave arises in the interaction region, when both wave components, described by the coefficients $A_1$ and $A_2$, interfere so that the resulting wave function does not extend outside the vicinity of the interaction region.

Until now, we have considered the quasi-bound states with zero angular quantum number, since they have the highest binding energy. However, when the interaction radius is not very small, the states with $m=2, 4, \dots$ also arise. The binding energy of these states, defined as $\Delta\varepsilon=\varepsilon_m-1$, decreases rather rapidly with increasing $m$. So, numerical estimations for the model parameters $a=0.6$, $v_0=0.35$ and $r_0=6$ show that the ratio of $\Delta\varepsilon$ at $m=2$ to that at $m=0$ is approximately 0.86. The width of the resonance, $\gamma$, on the contrary, increases with increasing $m$, so that the ratio of $\gamma$ at $m=2$ to $\gamma$ at $m=0$ is about 3.5. General tendency to widening the resonances with increasing $m$ comes from strengthening the hybridization. Indeed, in the BHZ Hamiltonian, the hybridization is determined by the quantities $a\hat{k}_{\pm}$ which increase with $m$, since $\hat{k}_{\pm}=-i\exp{(\pm i\varphi)}(\partial_r\mp m/r)$ for angular harmonics.

\section{Discussion and Conclusion}\label{S:Summary}

We have studied two-particle quasi-bound states of interacting electrons for a wide class of 2D materials with MHD, in which the Mexican-hat shape of dispersion arises due to the hybridization of the electron-like and hole-like bands that form the band spectrum. The key point in the theory of these states, as we have established, is the fact that the spinor structure of two-particle states must correspond to the composition of the atomic orbitals that form the band spectrum. The point is that the e-e interaction substantially changes the relative contribution of different atomic orbitals to the two-particle state formed due to the interaction, and therefore the effective reduced mass of electrons also changes. Only in the limit of weak interaction does it approach the mass determined by the band spectrum. 

The main result is that quasi-bound states are formed with energies above the top of the MHD, near which the effective reduced mass is negative. The properties of quasi-bound states are largely determined by the hybridization parameter $a$ of the electron and hole bands. Of greatest interest is the situation when $a\ll 1$ and the shape of the Mexican hat is strongly pronounced. In this case, the resonance width $\gamma$ of the local density of states near the e-e interaction region is extremely small, its dependence on $a$ is described as $\gamma\sim a^4$. At realistically small values of $a\sim 10^{-1}$, the quality factor of the resonance is about $10^5$, and the lifetime of the quasi-bound state is estimated on the level of $10^{-7}$\,sec for the material with the mass term $|M|=10^{-2}$\,eV.

An interesting property of the quasi-bound states, arising due to the presence of pseudospin degrees of freedom, is the fact that, under certain conditions, a quasi-bound state loses its connection with the continuum of band states and transforms into a BIC. This occurs at certain values of the amplitude of the e-e interaction potential and the radius of interaction. The physical mechanism for the formation of such states is due to two factors. First, an important role is played by the presence of different atomic orbitals whose wave functions are orthogonal in the absence of hybridization. Secondly, for the occurrence of a BIC, it is necessary to fulfill certain conditions for the interference of two components of the wave function, which arise due to the presence of different atomic orbitals in the interaction region. This mechanism, as far as we know, has not been discussed in the literature on BICs~\cite{hsu2016bound}. It differs significantly from the previously studied mechanism of BICs in 2D topological insulators caused by the destructive interference of two resonances of edge states coupled with a closely spaced nonmagnetic defect~\cite{SABLIKOV20151775}.

The spin structure of quasi-bound states can correspond to both a singlet state with $S_z=0$ and two triplet states with $S_z=\pm 1$, and the triplets differ noticeably in energy from the singlet. A singlet state with a given angular quantum number has a higher binding energy than the corresponding triplet state. In this case, the width of the singlet resonances is much smaller than that of the triplet ones.

Rotation of electrons as such is not necessary for the formation of quasi-bound states. The states with a larger angular quantum number have a lower binding energy. However, rotational motion is present, since it is intrinsic to the helical nature of the quasi-bound states, as well as to topological states in general.

The idea that long-lived quasi-bound pairs of electrons can arise due to the presence of a negative effective mass in some regions of the Brillouin zone stems from theoretical searches for the mechanisms of high-temperature superconductivity~\cite{Belyavskii2000}. In recent years it was studied mainly for graphene~\cite{PhysRevB.92.085409}. The situation of a MHD in bilayer graphene was considered recently within the frame of a single-band approach~\cite{kochelap2022rotating}. Since the equations of this approach are determined only by dispersion in one band, we can use the results of this study with the parameters of the BHZ model to understand which properties of quasi-bound states are lost in the single band approximation. Of course, it is clear that the spin effects are strongly distorted in this case, but such general properties as the energy and width of the resonance and the role of rotational motion, apparently, do not change, at least at a qualitative level. In this way we conclude that the single-band approximation gives incorrect results for the resonance width, which is of most interest. In the single-band approximation the width is significantly larger. Moreover, the single-band approach leads to an incorrect dependence of the resonance width on the angular number $m$ and, consequently, to an inadequate understanding of the role of rotation.

In this connection, we note that the conclusion that, in the presence of pseudospin degrees of freedom, the correct description of quasi-bound two-particle states goes beyond the framework of the single-band approximation is consistent with what is known. The problem should be solved on the basis of the spinor representation of wave functions taking into account these degrees of freedom. This is precisely how the theory of quasi-bound states in Dirac semimetals is constructed~\cite{PhysRevB.81.045428,PhysRevB.86.035425,PhysRevB.92.085409,downing2017bielectron}.

Possible manifestations of the quasi-bound pairs in experiments can be associated primarily with electron transport, since we can expect that the presence of quasi-bound states will lead to spin-dependent resonances of the e–e scattering, quite similarly to resonances of the single-particle scattering from a repulsive impurity~\cite{SABLIKOV2023115492}.In an electron system with a Mexican-hat dispersion, it can be expected that e-e scattering, which is greatly enhanced near the resonances, has a significant effect on the conductivity. This is due to the fact that the strong non-parabolicity and even more so the presence of several allowed values of the momentum at a given energy lead to the fact that the quasi-momentum may not be conserved in e-e collisions. It is known that in systems of this kind with broken Galilean invariance, e-e collisions lead to a significant change in the conductivity, and especially to a change in its dependence on temperature~\cite{PhysRevLett.106.106403,pal2012resistivity}. But the most interesting effects can be expected from the spin-dependent asymmetry of the e-e scattering, which can lead to the appearance of the spin and valley Hall effects~\cite{PhysRevB.106.235305}. Clarification of these issues can be the subject of further study. 

\begin{acknowledgments}
  This work was carried out in the framework of the state task for the Kotelnikov Institute of Radio Engineering and Electronics.
\end{acknowledgments}
        
\bibliography{2e_QBS}

\end{document}